\documentstyle[aps,prl,twocolumn,psfig]{revtex}
\begin{document}
\draft                       


\title{Computer investigation of the \\ energy landscape of amorphous silica}                           

\author{Philippe Jund and R\'emi Jullien}

\address{Laboratoire des Verres - Universit\'e Montpellier 2\\
         Place E. Bataillon Case 069, 34095 Montpellier France
	}


\maketitle


\begin{abstract}
The multidimensional topography of the collective potential energy function
of a so-called strong glass former (silica) is analyzed by means of classical molecular
dynamics calculations. Features qualitatively similar
to those of fragile glasses are recovered at high temperatures : 
in particular an 
intrinsic characteristic temperature $T_c\simeq 3500$K 
is evidenced above which the system starts to investigate non-harmonic 
potential energy basins.
It is shown that the anharmonicities are essentially characterized by a 
roughness appearing in the potential energy valleys explored by the system 
for temperatures above $T_c$.

\end{abstract}

\pacs{PACS numbers: 61.43.Fs, 61.20.Ja, 64.70.Pf}


\narrowtext

Even though the manufacturing of glasses by a quench from the high 
temperature liquid phase is a standard practice, the precise understanding of 
what happens at a molecular level in these
materials, when lowering the temperature through the ``glass transition'', 
remains an important theoretical
challenge. Several approaches have emerged in the literature depending whether 
the structural or the dynamical aspects are emphasized. In particular the use 
of the concept of topological frustration \cite{topo,jj99}
or the use of the mode-coupling theory \cite{mct} proceeds
 from completely different points of view. A promising way to
conciliate these approaches is the analysis of the shape of the potential 
energy function $\Phi (\overrightarrow
{r_1}, \overrightarrow{r_2},...)$ in
the multi-dimensional space of the coordinates of the interacting particles 
\cite{land,still2}. 
It has been recently shown \cite{sastry} that the topological
analysis of the
energy landscape of a  fragile glass-forming liquid described by a two-body 
Lennard-Jones (LJ) potential allows to explain
the distinct dynamical regimes experimentally observed.

In this letter, we present for the first time such an energy landscape analysis
in the case of a so-called strong glass former, namely vitreous silica, by 
means of classical molecular-dynamics (MD) calculations. As in the case of 
fragile glasses, from a quantitative analysis
of the inherent potential energy basins explored at various temperatures, 
one can define an {\em intrinsic} characteristic temperature, here
equal to $T_c\simeq 3500$K, above which the system experiences anharmonicities. 
These findings are in agreement with a recent simulation study in which a 
change in the dynamical properties of a similar liquid silica system 
has been observed around this temperature \cite{kob}, suggesting that 
supercooled liquid silica behaves like a fragile glass former at temperatures 
above  $T_c$.
Here we show that the anharmonicities
exhibited by the energy basins explored at temperatures above $T_c$ are 
essentially due to a roughness in the shape of the basins, 
as previously  observed in simpler liquid systems\cite{sastry,still}.
Using an original procedure, we perform a quantitative analysis
of this roughness and show that it is characterized by a typical
length (in the multidimensional configuration space) of about 0.2\AA\  
which is the signature of the annihilation of structural defects along 
the path down the potential energy valleys.

Our silica system consists of  216 silicon and 432 oxygen atoms confined in a cubic box of edge length $L=21.48$\AA\ 
which corresponds to a mass density very close to the experimental value of vitreous silica. 
The constant energy, constant volume
classical MD calculations have been performed  using the sophisticated potential first introduced by van 
Beest et al.  \cite{vanbest}
and justified by {\it ab initio} calculations. 
It has been recently shown to describe quite well structural \cite{jj99,vollmayr},
and vibrational \cite{taraskin}, as well as relaxational \cite{kob} and thermal \cite{jjpr}
properties of both, supercooled viscous liquid and glassy silica.
We have used the same input parameters and the same {\it modus operandi} as in our previous structural study \cite{jj99}.
After a full equilibration of the liquid (about 28 ps) the system has been cooled down 
to zero temperature at a quench rate
of $2.3\times10^{14}$ Ks$^{-1}$ which has been obtained by removing the corresponding 
amount of energy from the total energy
of the system at each iteration. At several temperatures during the quenching process the 
configurations (positions
and velocities) have been saved. 
With the use of these configurations as inputs of the MD calculations, the system has 
then been 
allowed to relax for a maximum duration of 84 ps after the quench (during 
this constant volume relaxation period the temperature
of the system is only slightly increasing).
 The same procedure has been repeated for ten different initial liquid configurations and consequently 
all the physical quantities
reported below always result from an average over these ten independent samples.

For each collected temperature, immediately after the quench or after a 
given relaxation time (42 or 84 ps), the typical potential energy basin 
sampled by the system has been 
investigated using  a procedure described by Della Valle and Andersen 
\cite{della} which allows to perform a steepest descent from the initial 
multicomponent space configuration
down to the closest underlying potential energy minimum, often called the 
``inherent structure'' in the literature \cite{still2}.
For that purpose a modified version of the MD algorithm is adopted.
At each MD step, the scalar product of the velocity by  the force is  
calculated for each particle.
If the product is positive the velocity of the particle is replaced by its 
projection on the force and otherwise the velocity is set to zero. During 
the descent, both the 
distance and the potential energy $\Phi$ are calculated, allowing to 
determine precisely the shape of the energy basin. There are different ways 
to define the distance per particle in the
multiparticle space. We have chosen to calculate either
the direct distance $x$ from the initial configuration, defined by 
\cite{remarque} :
$$x = \sqrt{ {\sum_i m_i (\overrightarrow{r_i} -\overrightarrow{r_i}^{(0)})^2 \over \sum_i m_i} }$$
where the sum runs over all the particles located at $\overrightarrow{r_i}$ 
with a mass $m_i$, or the cumulated distance, $X$, calculated along the 
steepest descent path by :
$$X   = \sum_n \sqrt{\sum_i m_i (\overrightarrow{r_i}^{(n)} -\overrightarrow{r_i}^{(n-1)})^2 \over \sum_i m_i}$$
where $n$ labels the MD steps.
The process is stopped when $\Phi$, which is a monotonically decreasing 
function of $x$ (resp. $X$), reaches a minimum $\Phi_m$ at $x=x_m$ 
(resp. $X=X_m$). 
Practically we have chosen to stop the descent when 
$\Phi^{(n)}-\Phi^{(n+1)} < 10^{-6}$ eV.

As a first result we present in Fig.1, the potential energy minima 
$\Phi_m$ obtained at different temperatures. The curve obtained immediately
after the quench is similar to the one obtained previously 
with a different potential \cite{della}. Of course, the system
investigates energy basins of lower minima at lower  temperatures. 
The slowing down of the decrease of $\Phi_m$ with decreasing temperature
occurring  between 4000K and 3000K is the signature of the glass transition 
as the system gets trapped after the quench in energy basins with almost the
same minimum below a given temperature $T_g$. This temperature is consistent 
with the estimate $T_g\simeq 3500$K obtained from a structural analysis done 
for the same system with the same quenching rate \cite{jj99} but also with 
the extrapolated value obtained in a different study concerning the influence 
of the quenching rate on the properties of the system \cite{vollmayr}.
Moreover, in agreement with the study done on the LJ glass \cite{sastry}, 
one observes in Fig.1 that the inherent structure depends on the duration of
the relaxation process after the quench. The relaxed curves exhibit a 
minimum around $T_g$ which is more and more marked as the aging time 
increases. This can be easily understood since  
for temperatures close to $T_g$ the system takes advantage 
of the relaxation process in order to find lower energy minima, because it 
has enough kinetic energy to have a chance
to cross the energy barriers between minima. Of course this chance  
becomes considerably smaller at lower
temperatures in agreement with what is known from the thermal evolution of 
the relaxation time in the
glassy phase of silica (note that in \cite{sastry} the energy per particle at 
$T=0$K depends on the cooling rate while we do not observe significant 
relaxation effects at $T=0$K). These results are consistent with a previous 
analysis of history effects in the same system \cite{jj00} 
and also with the fact that $T_g$ should be smaller for a lower quenching 
rate \cite{vollmayr}.
If we take into account both, the 15 ps necessary to reach 3500K from 7000K 
and the 84 ps of further relaxation at 3500K, we obtain an effective 
quenching rate about six times smaller than the one we have used. According 
to the dependence of $T_g$ with the quenching rate proposed by Vollmayr 
et al. \cite{vollmayr}, this would correspond to a decrease of $T_g$ by 
more than 500K.

In Fig.2 we have plotted, as a function of temperature, $x_m^2$ 
the square of the distance between a given initial configuration and the 
corresponding position of the inherent structure. This curve is remarkably 
similar to what has been previously obtained for
Lennard-Jones glasses \cite{sastry} and can be interpreted in the same manner. 
The squared distance $x_m^2$ is linear with temperature up to a characteristic 
temperature $T_c\simeq 3500$K (here very close to $T_g$) above which 
anharmonicities appear. 
The same analysis as the one done
in \cite{sastry} shows that $T_c$ corresponds to
a   change in the nature of the relaxation process, i.e. from diffusion 
above $T_c$ to hopping below $T_c$.
This is consistent with the recent work of Horbach and Kob on the same 
system \cite{kob} who found a breakdown of the Arrhenius behavior of the 
transport coefficients above a temperature close to 3300K and suggested 
that this characteristic temperature corresponds to the $T_c$ predicted by 
the mode coupling theory.                                                       
A more striking result is that the curves in Fig.2, in contrast with those 
in Fig.1, are not very sensitive to the duration of the relaxation process 
after the quench. This is also true for another characteristic quantity 
of the energy basins leading to the inherent structures, 
namely $\Delta\Phi = \Phi^{(0)}-\Phi_m$, the energy difference between the
initial structure and the inherent structure, which has been plotted versus 
$T$ in the inset of Fig.2. It exhibits the same departure from the expected
harmonic linear regime ($\Delta\Phi = \frac{3}{2}k_{B}T$ represented by 
the dashed line in the inset) at $T_c$ and also the same remarkable 
stability against relaxation. 
This shows that as the aging time increases the system explores deeper and 
deeper basins (especially around $T_g$) as shown by the variation 
of $\Phi_m$ in Fig.1 but the 
intrinsic characteristics of these basins, the mean width and the mean height 
(associated with  $x_m^2$ and $\Phi^{(0)}-\Phi_m$ resp.), do not depend on their 
absolute position $\Phi_m$ in the energy scale. Therefore the characteristic 
temperature that we can define here, $T_c\simeq 3500$K, is intrinsic and 
does not depend on the quenching rate. The fact that we obtain $T_g$ close 
to $T_c$ is simply due to our very large quenching rate.
It is worth noticing that the same 
qualitative behavior than the one observed in Fig.2 is also obtained by 
plotting $X_m^2$ the 
square of the cumulated distance instead of $x_m^2$. In that case  the departure from a linear regime above $T_c$ is even more pronounced as we find that the ratio $X_m/x_m$ remains constant (of order 1.2) for $T<T_c$
and increases markedly with temperature for $T>T_c$.

To investigate further the nature of the anharmonicities of the energy basins explored by the system for temperatures above $T_c$, we propose a 
quantitative analysis of the curves $F(X)=-d\Phi/dX$ where $X$ is the above 
defined cumulated distance (we have used $-d\Phi/dX$ instead of $-d\Phi/dx$ 
because it corresponds better to a local characteristic of the shape of the 
basins). Typical examples of such curves are given in Fig.3
for samples relaxed during 84 ps after the quench (as already shown in Fig.2 
the aging time does not influence significantly the numerical results). 
It turns out that for $T>T_c$, the  $\Phi(X)$ curves, 
while always decreasing, exhibit step-like  singularities 
evidenced by clearly visible peaks in the derivative. 
Note that another manifestation of what can be called a roughness 
is the increase with 
temperature of the ratio $X_m/x_m$ above $T_c$, mentioned earlier (the          
same observation has already been done for LJ glasses \cite{sastry}). 
To analyze quantitatively the roughness of  the $F(X)$ curves, we
have first eliminated the overall mean general evolution by calculating the 
difference $f(X) = F(X)-<F(X)>$ where $<F(X)>$
is a local average of the data between $X-\delta X$ and $X+\delta X$ (for 
convenience we have chosen $\delta X= X_m/20$ and limited the range of $X$ 
values between $4\delta X$ and $X_m-\delta X$). The $f(X)$
curves corresponding to the $F(X)$ curves depicted in Fig.3 are shown in the 
inset of the figure (they have been artificially shifted in the vertical 
direction for clarity).
Subsequently the roughness  of the curves $f(X)$ has been analyzed by 
following a standard method \cite{baraba}  
which consists in calculating the power spectrum $S(k)$ defined as the 
Fourier transform of the autocorrelation product $g(\xi)$ = $<f(X+\xi)f(X)>$ 
where the average is performed not only over the
$X$ values but also over  ten independent samples. The results of this 
analysis are reported in Fig.4.  
In this figure we have reported as a function of
temperature the mean intensity of the peaks measured
by the standard deviation $\sigma_f$ of the $f(X)$ curves, which is equal to 
the square root of $g(0)$. Despite the error bars
the curve exhibits a characteristic
sigmoidal shape, indicating that the roughness only exists in the higher 
energy basins explored
for $T>T_c$ and that the mean intensity of the peaks seems to saturate at 
very high temperatures. 
Furthermore, for $T>T_c$, we observe that the power spectrum goes through a 
maximum and decays like $k^{-1}$ after this maximum. This means that there 
exists a typical length (the inverse of the location of the
maximum) $X_r$ characteristic of the mean distance between successive peaks 
while the curve between two
peaks can  practically be considered as ``smooth''. For $T<T_c$ there is no
visible maximum in the power spectrum which behaves roughly as $k^{-1}$ over 
the whole $k$ range. This is a further indication that the very weak roughness 
in the basins explored below $T_c$  has no significance and that the low 
temperature basins can be considered as smooth. 
The estimated values for $X_r$  above $T_c$ have been reported in Fig.4.
This typical length increases slightly with $T$ from about 0.15 \AA\ at 4000K to about 0.23 \AA\ at 7000K.
It is interesting to relate the typical distance $X_r$ in the 
multidimensional configuration space to peculiar structural rearrangements 
occurring during the down-hill potential energy minimization process.
On some specific high temperature samples we have compared the 
configurations between two successive peaks in $F(X)$ and we have in each case 
observed the elimination of a single specific defect like a triconnected 
silicon atom or edge-sharing tetrahedra etc... In such rearrangements a 
"perturbed" cluster of about 30 to 50 {\em connected} atoms, is observed. 
Generally the largest displacement of about 0.5 to 
0.7 \AA\ is observed for an oxygen atom at (or very close to) the defect.
Therefore the typical value of $X_r$ results from an average between the 
largest displacement near the defect and the ``screening cloud'' of displaced 
atoms connected to it. One can understand that $X_r$ becomes insignificant
below $T_c$ because the defects become rare in the low temperature basins
as already shown in \cite{della}.

In conclusion we have numerically investigated the potential energy landscape 
of super cooled liquid silica described by the BKS potential using a 
steepest-descent molecular-dynamics scheme. We have shown that the inherent 
structures sampled depend strongly on the effective cooling-rate especially 
around $T_g$ similarly to what was found in a Lennard-Jones glass \cite{sastry}. Nevertheless at a given temperature the characteristics of the
energy basins (mean height and mean width) seem to be insensitive to the 
history of the system. From a quantitative analysis of the potential 
energy valleys explored at various temperatures
we have evidenced a characteristic temperature $T_c$ above which 
non-harmonic effects become dominant.  
This is consistent with an other recent study \cite{kob} and indicates that 
strong and fragile glass formers are quite similar when studied near $T_c$ 
where a change in the nature of the relaxation process takes place in the 
liquid phase. We think that the distinct characteristics of a strong glass 
former appear mainly in the supercooled liquid and glassy phases below $T_c$. 
Unfortunately the high value of $T_g$ that we obtain in our MD simulations 
does not allow us to study the temperature range between $T_g$ and $T_c$.
Furthermore using an original quantitative analysis, we have shown that the 
anharmonic character of the higher energy valleys explored by the system 
above $T_c$, is due to some roughness occurring in the shape of the potential 
energy basins. The existence of such  roughness  has already been invoked in 
the case of simpler systems \cite{sastry,still} but not quantitatively analyzed. In the case of silica  we have shown that this roughness is characterized
by a typical length of about 0.2 \AA\  in the multi-dimensional configuration 
space, and can be associated with a sequential elimination of defects when 
following the path leading down to the inherent structures.

Part of the numerical calculations were done at CNUSC
(Centre National Universitaire Sud de Calcul), Montpellier.


\vspace*{-0.5cm}
%
\newpage

\begin{figure}
\psfig{file=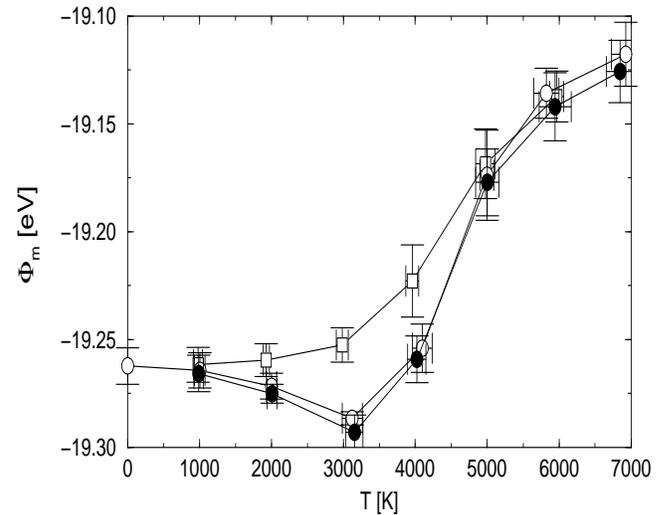,width=8.5cm,height=7cm}
\caption{
Plot of $\Phi_m$, the minimum value of the potential energy basin 
explored by the system, as a function of the temperature $T$  down to which 
the system has been cooled from the liquid state.
The three curves correspond to the different durations of the relaxation period
 after the quench: $\Box$: no relaxation; $\circ$: 42ps; $\bullet$: 84 ps. 
The data result from an average over ten independent samples.
}
\label{Fig1}
\end{figure}

\vspace*{-0.6cm}
\begin{figure}
\psfig{file=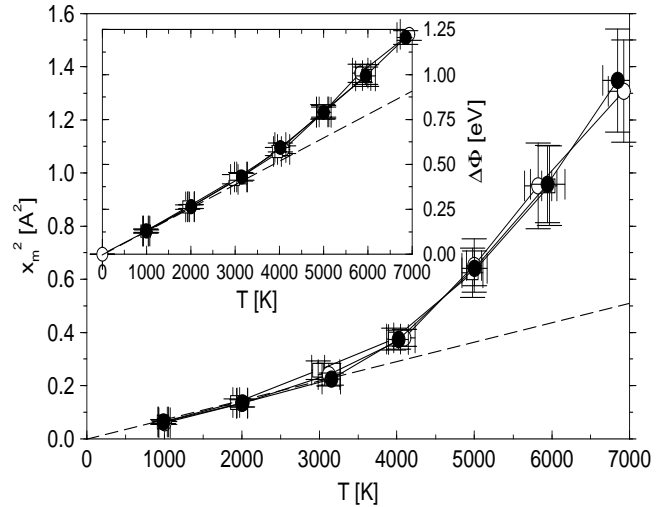,width=8.5cm,height=8cm}
\caption{
Plot of $x_m^2$, the mean square distance between the initial 
configuration and the corresponding minimum potential location as a 
function of temperature.  The three curves correspond to the different 
durations of the relaxation period  after the quench:\\ $\Box$: no 
relaxation; $\circ$: 42ps; 
$\bullet$: 84 ps. The data result from an average over ten independent samples. The straight line corresponds to a pure linear behavior with temperature. \\ 
In the inset the variation
of $\Delta\Phi = \Phi^{(0)}-\Phi_m$, the energy difference between the 
initial configuration and the inherent structure, is plotted versus temperature 
with the same conventions (the dashed line represents the harmonic linear regime: $\Delta\Phi = \frac{3}{2}k_{B}T$).
}
\label{Fig2}
\end{figure}

\begin{figure}
\psfig{file=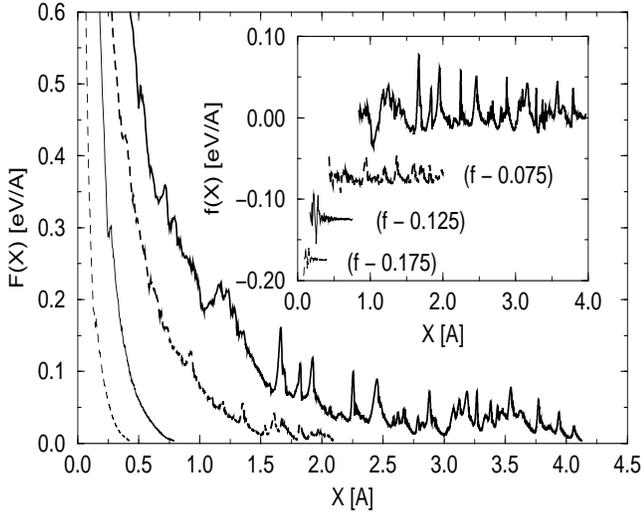,width=8.5cm,height=7cm}
\caption{
Plot of the derivative $F(X)=-{d\Phi\over dX}$ of some typical 
energy basins explored at different temperatures in the case of a sample 
relaxed 84 ps after the quench. From left to right the curves correspond 
to $T=1000, 3000, 5000$ and 7000 K, 
respectively. In inset are  shown  the corresponding curves $f(X)$ 
obtained by making the difference between F(X) and a local average 
as explained in text. 
In the inset the curves have been arbitrarily shifted vertically for clarity.
}
\label{Fig3}
\end{figure}

\begin{figure}
\psfig{file=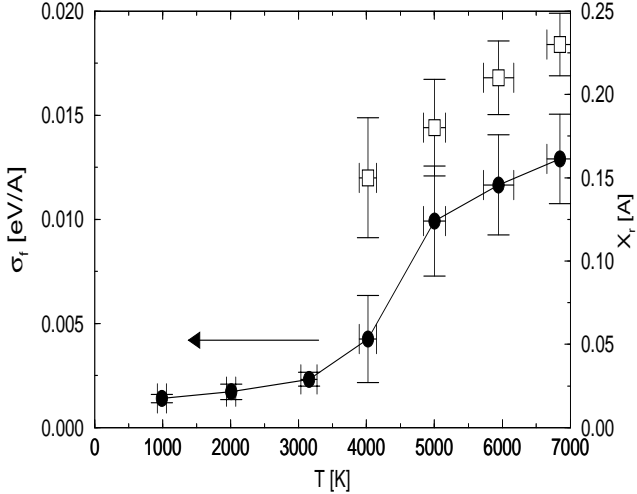,width=8.5cm,height=7cm}
\caption{
Plot of the mean-square intensity $\sigma_f$ ($\bullet$) of the 
roughness in the
space-derivative of the potential curve as a function of 
temperature after averaging over ten samples with a relaxation period of 
84 ps after the quench.
On the same figure the typical distance $X_r$ ($\Box$) between 
the peaks in $f(X)$ is plotted as a function of temperature
(right vertical axis).
}
\label{Fig4}
\end{figure}


\begin{references}
\bibitem{remarque} Note that, in contrast to other studies, we have 
weighted the sum  by the mass of the particles in the definition of $x$. On a few examples, we have checked that the results are not depending on the definition of $x$.
\end{references}
\end{document}